\documentstyle[12pt,psfig,scipp,equations]{article}
 
\def\bra#1{\left\langle #1\right|}
\def\ket#1{\left| #1\right\rangle}
\makeatletter
\newcount\@tempcntc
\def\@citex[#1]#2{\if@filesw\immediate\write\@auxout{\string\citation{#2}}\fi
  \@tempcnta\z@\@tempcntb\m@ne\def\@citea{}\@cite{\@for\@citeb:=#2\do
    {\@ifundefined
       {b@\@citeb}{\@citeo\@tempcntb\m@ne\@citea\def\@citea{,}{\bf ?}\@warning
       {Citation `\@citeb' on page \thepage \space undefined}}%
    {\setbox\z@\hbox{\global\@tempcntc0\csname b@\@citeb\endcsname\relax}%
     \ifnum\@tempcntc=\z@ \@citeo\@tempcntb\m@ne
       \@citea\def\@citea{,}\hbox{\csname b@\@citeb\endcsname}%
     \else
      \advance\@tempcntb\@ne
      \ifnum\@tempcntb=\@tempcntc
      \else\advance\@tempcntb\m@ne\@citeo
      \@tempcnta\@tempcntc\@tempcntb\@tempcntc\fi\fi}}\@citeo}{#1}}
\def\@citeo{\ifnum\@tempcnta>\@tempcntb\else\@citea\def\@citea{,}%
  \ifnum\@tempcnta=\@tempcntb\the\@tempcnta\else
   {\advance\@tempcnta\@ne\ifnum\@tempcnta=\@tempcntb \else \def\@citea{--}\fi
    \advance\@tempcnta\m@ne\the\@tempcnta\@citea\the\@tempcntb}\fi\fi}
\makeatother
\begin{document}

 \rightline{SCIPP 96/41}
 \rightline{BUTP 96/25}
 \rightline{hep-ph/9611353}
 \rightline{September 1996}
 \vskip4pc
\begin{center}
{\large NEUTRINOLESS DOUBLE BETA DECAY AND ITS ``INVERSE"} \\
                                                                                
\vspace*{1pc}
{CLEMENS A. HEUSCH}                                        \\
{\sl Santa Cruz Institute for Particle Physics}            \\
{\sl University of California, Santa Cruz}                 \\
\vspace*{6pt}
{PETER MINKOWSKI}                                          \\
{\sl Institute for Theoretical Physics}                    \\
{\sl University of Bern, Switzerland}                      \\
\end{center}
\vspace*{1pc}

{\begin{center}
\begin{minipage}{5truein}
\small
\centerline{ABSTRACT}
\vskip3pt\noindent
Recent considerations by these authors pointed out the attractive               
features which a search for the exchange of heavy Majorana neutrinos            
could have for solving the mass and the lepton number puzzles for
all neutrinos, in TeV-level electron-electron scattering. In the present        
note, we show that, contrary to subsequently published arguments,               
non-observation of neutrinoless double beta decay has, to date, no              
bearing on the promise of this important task for future linear electron        
colliders.                                                                      
\end{minipage}\end{center}
}
                                                                                
\par\vspace*{2pc}  \normalsize
Recent developments in the planning for electron colliders in the TeV           
energy region have renewed interest in the possibility to investigate           
the reaction                                                                    
\begin{equation}
e^-e^- \to W^-W^-,        
\label{eqone}
\end{equation}
which can proceed by means of the exchange of a Majorana neutrino.
While previous work on this reaction (Refs.~\cite{ushio,rizzo,%
londonetal,maalampi}) 
had not led to
promising experimental prospects, we showed that the new generation of          
presently projected linear colliders can in fact deliver luminosities           
compatible with the production of convincing signals for reaction 
(\ref{eqone}), or, failing that, important new limits on the masses
and couplings of heavy Majorana neutrinos.
This is predicated on the availability of highly polarized electron             
beams at center-of-mass energies upward of 500 GeV, where left-handed           
electrons will be able to interact {\it via} the exchange of heavy left-handed        
Majorana singlet neutrino states with masses $\ge 1$ TeV. For the case
of two longitudinal $W^-$ in the final state, the relevant
cross-section increases $\sim s^2$ in the kinematic region 
$2m_W < \sqrt s < m_N$. The existence of two or more such singlets 
follows naturally from
SO(10) decomposition, and could be the key ingredient for our                   
understanding of both the observed very light masses of the three known         
neutrino states, and of the meaning of lepton number and its                    
conservation or non-conservation.                                               
                                                                                
Several authors \cite{belanger,pantis} 
have argued that our calculations cannot serve as         
the basis for a possible successful observation of reaction (1): they           
maintain that existing limits on the related process involving                  
neutrinoless double beta decay [hereafter  $\beta\beta_{0\nu}$]
already exclude it. In so doing, the authors of Ref.~\cite{belanger} 
explicitly refer
to reaction (\ref{eqone}) 
as ``inverse neutrinoless double beta decay".

The fallacy of their argument can, in a nutshell, be gleaned from this          
misnomer: in the present note, we show that only a profound misreading          
of the $\beta\beta_{0\nu}$ reaction mediated by heavy Majorana neutrinos
can lead to the conclusions of Refs.~\cite{belanger,pantis}. 
Let us consider the diagrams
in Fig.~1: we ask ourselves whether the lack of observation of
graph 1b can serve to impose tight constraints on the observability of          
graph 1a.

\begin{figure}[htbp]
\centerline{\psfig{file=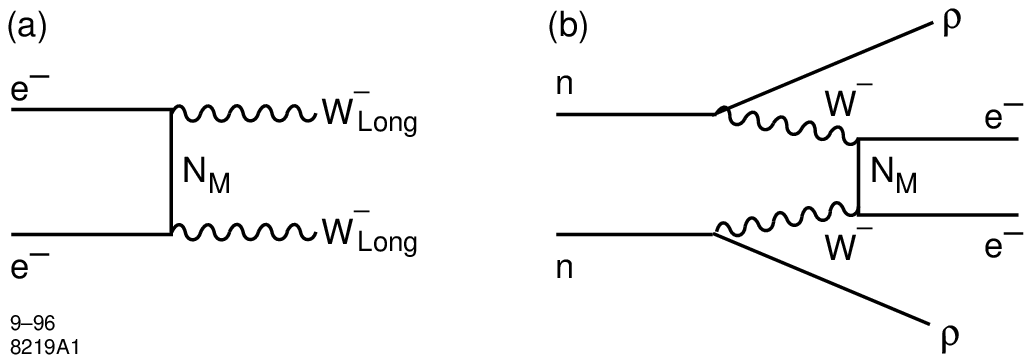,height=3.5cm}}
\fcaption{
a) diagram for the quasi-elastic production of two longitudinal $W^-$
in the scattering of TeV-level left-handed electrons, mediated by               
massive Majorana neutrino exchange.                                             
b) Neutrinoless double beta decay {\it via} exchange of a heavy Majorana
neutrino, with the hadronic part seen at the nucleon level.                     
}
\label{fig:one}
\end{figure}
                                                                                
The best present evidence on the non-observation of the $\beta\beta_{0\nu}$
reaction is from the experimental limit on the process                          
\begin{equation}
^{76}Ge \to ^{76}Se\ e^-e^-,        
\label{eqtwo}
\end{equation}
with $\tau_{1/2} > 5 \times10^{24}$ years. How do we translate this into a
limit on reaction (\ref{eqone})?  
                                                                                
Literally, the inverse to reaction (\ref{eqtwo}) 
is
\begin{subequations}
  \begin{equation}
  e^-e^-\ ^{76}Se \to\ ^{76}Ge \label{eqthree-a} \\      
 \end{equation}
{\hbox{or, somewhat less rigidly,}}
 \begin{equation}
  e^-e^- \to ^{76}Ge\ ^{76}\overline{Se}. \label{eqthree-b}\\       
  \end{equation}
\end{subequations}
Neither of these two reactions are experimentally realizable, but the           
argument points up a troubling question: Can the (sub-)reactions that
lead to the decays $nn \to ppe^-e^-$ occur freely inside the nuclei $^{76}$Ge and
$^{76}$Se?                                                                           
                                                                                
At this point, we have to remark that graph 1b symbolizes a process that        
acts over distances of order $(m_W)^{-1}$ or about $10^{-16}$cm. If, on the
other hand, the exchanged neutrinos had the masses of the known light           
varieties, $m_\nu <1$ eV, the range over which the interaction extends would
be $>10^{11}$ times larger. In the heavy ($\sim$1 TeV) neutrino exchange case,
the subprocess we have to study is                                              
\begin{eqnarray}
dd &\to& uue^-e^- , \nonumber   \\   
(nn)& &(pp) \\
((^{76}Ge))& &  ((^{76}Se)) \nonumber
\end{eqnarray}
where the symbols in single and double parentheses stand for the                
constraining configurations within which  the interacting particles          
of the lines above have to be considered, respectively: the hadronic            
systems must be treated on the quark level, but are heavily constrained         
by nucleon-nucleon forces, and the latter are in turn constrained by the        
specific wave functions of their host nuclei. It is worth noting that,          
in this context, the study of $\beta\beta_{0\nu}$ {\it via} heavy $N$ exchange
can be
seen as a unique probing of nuclear structure on the quark correlation          
level, at distances $<10^{-16}$cm \cite{heusch}. 
What then is our chance of observing
$dd$ overlap at these distances, within the constraints of eq.~(4)? Let us
first determine the constraints we can easily establish:                        
                                                                                
\begin{enumerate}
\item The final-state electrons have to emerge in an overall $S$ state,
as a spin
singlet. This, in turn, imposes a spin singlet configuration on the dd          
wave function.                                                                  
                                                                                
\item  To achieve an overall antisymmetric wave function for the $l=0\ dd$
system, the product of space and SU3$_c$ wave functions has to be
symmetrical. This leaves only the 6 representation of SU3$_c$, imposing a
suppression factor of 2/3.                                                      
                                                                                
\item To evaluate the strong Hamiltonian density involved in the $dd \to
uue^-e^-$ subgraphs of Fig.~2a, we have to keep the constraints imposed by
the surrounding nuclear and nucleon environment in mind, as                     
schematically shown in Fig.~2b. This leads to two further suppression
factors to be determined: one is due to the color Coulomb repulsion of          
the $d$ quarks, the other to the collective pull which the saturated nucleon         
configurations of two neutrons exert on each quark that may be drawn into
an interaction with a quark from another nucleon.                               
                                                                                
\item Finally, the resulting Hamiltonian density operator will have to
include the leptonic weak current operator,  integrated over the
appropriate interaction volume, and then sandwiched between the mother          
and daughter nuclei's wave functions. 
\end{enumerate}

\begin{figure}[htb]
\centerline{\psfig{file=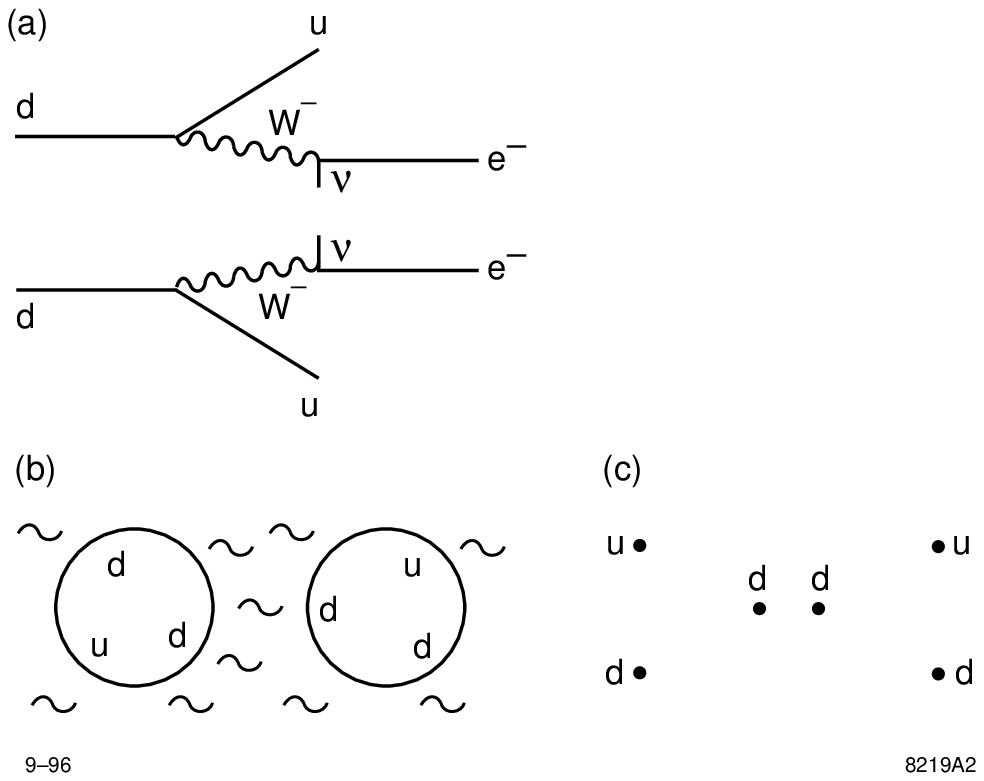,height=7.2cm}}
\fcaption{
a) Two quark-level sub-diagrams for the neutrinoless double beta
decay process of Fig.~1b, prior to the identification of the exchanged
neutrino in terms of a massive Majorana particle.                               
b) Schematic picture of the agents in neutrinoless double beta
decay: 6 quarks in two neutrons inside the nuclear environment.                 
c) Schematic arrangement of quarks in the process of a): the
color sextet interaction repels the two central $d$ quarks; the color
singlet interaction pulls the central $d$ quarks in opposite directions,
toward the CM of the remaining $ud$ pairs. See text.
}
\label{fig:two}
\end{figure}
                                                                                
Each of these effects will lead to a suppression factor. In an attempt          
to write down the Hamiltonian density of the highly local quark-quark           
Hamiltonian density in the overall expression
\begin{equation}
  H = G^2_F \left[ \bar e_\alpha \bar u^{c\alpha} d^c_\gamma
        (x_2) \bar e_\beta \bar u^{b\beta} d^{b\gamma}
        (x_1)\right] _{x_2\to x_1}                           
\end{equation}
we can write, using the customary lepton-hadron factorized expression,          
\begin{equation}
H \propto  U_{eN}^2 / m_N
  \left \lbrack \bar{e}_{\alpha} \bar{e}_{\beta} \right\rbrack\
   \left \lbrack \bar{u}^{c \alpha}
    d^{c \gamma}\bar{u}^{b \beta} d^{b}_{\gamma}
       \right\rbrack
\end{equation}
$U_{eN}$ is the mixing angle for electron/heavy neutrino $N$ with mass $M$,
the $b,c$ are color indices. The high degree of locality that
governs the interaction involving both sub-diagrams of Fig.~2a permits us        
to rewrite the hadronic Hamiltonian density in eq.~(6) such as to 
pair like-flavor quarks:
\begin{equation}
H_{q} (x) = \left[ \bar{u}^{b}_{\alpha}
            \bar{u}^{c \alpha} (x) \right]
            \ \left[ d^{c \gamma} d^{b}_{\gamma} \ (x)\right]\ .
\end{equation}

  This density operator can then conveniently be sandwiched between the
nuclear state vectors for mother $(A,Z)$ and daughter $(A,Z+2)$ nucleus, 
for a hadronic matrix element                                                       
\begin{equation}
\bra{A,Z+2;p_{2} } H_{q}  (x)  \ket{A,Z;p_{1}} =
  V^{-1} e^{i q x}  \varrho_{21}\ ; \quad
      q = p_2 - p_1 \ .
\end{equation}
                                                                                
$V$ is a normalizing volume, $\rho$ is a density matrix that
contains the
Fermi and Gamow-Teller structure of the interaction when expressed
in the current--current form (VV and AA on the
quark level). It can be saturated with a complete set of the
possible intermediate states with the requisite energy and                      
momentum \cite{noteone}.  
We will narrow our interest down to two-nucleon
correlations; they will dominate the small-distance behavior in the case        
of $N_M$ exchange. Recall that $H_q$ in eq.~(8) is a four-quark
operator: the $H_q$ operator does not ``see" the Ge, As, Se nuclei;
rather, its interaction involves the 76 nucleons in terms of
their 228 quarks.

We therefore have to try and evaluate the quark-quark suppression               
factors in the density matrix $\rho_{21}$ of eq.~(8). They are a function of the
relative distance $r_{12}$ of any two interacting quarks, at $r_{12}$ values in
the region of the ``hard-core", $r_{12}<0.3$fm.
First, there is the 2/3
factor due to the spin singlet requirement (see above). Second, the             
repulsive color sextet interaction can be reasonably estimated by a WKB
method for an evaluation of the color Coulomb barrier. A straightforward
relativistic treatment leads to a barrier penetration/inhibition factor
\begin{equation}
F_B = e^{- \pi\alpha_s/3},
\end{equation}
irrespective of the nuclear environment.                                        
                                                                                
Third, as Fig.~2b indicates, there are similar inhibition factors to be          
expected from the interaction of the remaining two quarks in the two            
interacting neutrons, two each that are not directly overlapping. We            
illustrate this schematically in Fig.~2c: each of the two central $d$
quarks is being ``pulled on" by a $u$ and a $d$ quark from its ``own" 
neutron, trying to keep the straying companion in the color singlet                      
configuration. Although the relevant Clebsch Gordan coefficient may make        
these inhibition factors somewhat stronger, we approximate them by a            
joint ansatz of                                                                 
\begin{equation}
      F_{nn} \approx F_B^2 = e^{-2/3 \pi \alpha_s}.
\end{equation}
This factor of order 1/9 is very conservatively estimated, given that           
the color force increases considerably above the $r^{-2}$ level known for the
small-distance behavior valid for the initial $dd$ interaction at $r_{12}$
values of $>~1/3$fm, for this somewhat longer-range color singlet
restoration force. We feel justified in regarding this suppression as           
amply supported by such evidence as the loose binding of the deuteron           
and the non-existence of bound $nn$ and $pp$ states. Note that this
repulsive vector interaction can also be modeled in terms of omega meson        
exchange between the two neutrons, leading to an inhibition factor              
stronger than the one resulting from eq.~(10).
                                                                                
Lastly, let us recall that the entire process thus inhibited by a factor        
bounded from above by $2/3 \times (1/3)^3 = 2/81$, or about 0.025, 
but likely
to be considerably smaller, has to happen inside the ``mother-daughter"
nuclear system. For the $^{76}$Ge $\beta\beta_{0\nu}$ decay with the best 
present
limits, this means that the above suppression modifies the nuclear wave         
function overlap symbolized by the projection operator                          
\begin{equation}
  \hat P \equiv \ket{^{76}Se^{-2u}} \bra{^{76}Ge^{-2d}} ,          
\end{equation}
which effectively sums over all intermediate states in the density matrix       
$\rho_{21}$ that contain two $d$ quarks less than $^{76}$Ge, 
two $u$ quarks less than $^{76}$Se \cite{PM}.
                                                                                
Finally, we compare our conservative estimate of the overall inhibition         
factor with the recent literature: it reduces the exclusion zone 
for
observable heavy Majorana neutrino masses from the  estimate made by
Pantis et al.~\cite{pantisetal} 
from $(1/m_N)_L^{-1} \gsim 6.7\times10^3$ TeV to
\begin{equation}
{(m_N)_L\over |U_{eN}|^2} = 0.025 \times 6.7 \times 10^3\ {\rm TeV}.
\end{equation}
With the mixing parameter $|U_{eN}|^2 = (2-40) 10^{-4}$ 
Ref.~\cite{heuschmink1}, 
this puts
\begin{equation}
(m_N)_L > \left\{ {0.67\ {\rm TeV}\atop 0.033\ {\rm TeV}}  \right\}\
         {\rm for\ the}\ \left\{ {{\rm upper}\atop {\rm lower}} \right\}
       \ {\rm limit\ on}\ |U_{eN}|^2 .
\end{equation}
Similarly, it moves the exclusion zone advocated in Ref. \cite{belanger} 
for the possible observation of          
process (1) in the face of existing evidence from neutrinoless double           
beta decay searches, as drawn in the $U_{eN}^2$ vs. $m_N$ plane, 
well out of danger's way \cite{notetwo}. 
                                                                                
We conclude that established limits on the observation of neutrinoless          
double beta decay do not in any way preclude the observability of               
process (1), and thereby the possible discovery of TeV-level Majorana           
masses in electron-electron scattering. The next generation of electron         
linear colliders thus has a highly attractive chance of unraveling the         
major mystery that shrouds our understanding of the observed lepton             
spectrum and forces an illogical treatment of the lepton sector on the
Standard Model.

%
%
\def\PRL#1&#2&#3&{\sl Phys.\ Rev.\ Lett.\ \bf #1\rm ,\ #2\ (19#3)}
\def\PRB#1&#2&#3&{\sl Phys.\ Rev.\ \bf #1\rm ,\ #2\ (19#3)}
\def\PRP#1&#2&#3&{\sl Phys.\ Rep.\ \bf #1\rm ,\ #2\ (19#3)}
\def\NPB#1&#2&#3&{\sl Nucl.\ Phys.\ \bf #1\rm ,\ #2\ (19#3)}
\def\PL#1&#2&#3&{\sl Phys.\ Lett.\ \bf #1\rm ,\ #2\ (19#3)}
%


\begin{thebibliography}{00}
\bibitem{notetwo}
 Note that Fig.~5 of Ref.~\cite{belanger} has to be corrected for the presence of a
high degree of polarization in both incident electron beams. With               
$P_e=90$\%, this raises the sensitivity by a factor 3.24 at the same time
that the lower limits to the exclusion zone are decreased by the factor         
0.025.
\end{thebibliography}
\end{document}